\documentstyle[12pt]{article}

\begin{document}

\begin{center}

{\large\bf Some Advantages of SUSY SU(4) \(\times\) SU(2)\(_ L \times\) 
SU(2)\(_R\) Model}\\[.2in]

{\large\bf in String Derived SO(10) GUTs }\\[1.0in]

{\bf  Akihiro Murayama}\footnote{
E-mail: murayama@ed.shizuoka.ac.jp} \\[0.2in]

Department of Physics, Faculty of Education, Shizuoka University \\ 
836 Ohya, Shizuoka 422, Japan  \\

\vglue 2.5in

{\bf abstract}

\end{center}

A {\it D}-parity violated SUSY SU(4) \(\times\) SU(2)\(_ L \times\)
SU(2)\(_R\) gauge model with the Higgs sector 
\(2\{(\mbox{\boldmath $4$},\mbox{\boldmath $1$},\mbox{\boldmath
$2$})+(\overline{\mbox{\boldmath $4$}},\mbox{\boldmath $1$}, \mbox{\boldmath 
$2$})\}+
(\mbox{\boldmath $1$},\mbox{\boldmath $2$},\mbox {\boldmath $2$})+\mbox{some}~
(\mbox{\boldmath $1$},\mbox{\boldmath $1$},\mbox {\boldmath $1$})\)'s is shown 
to have the following 
advantages: (i) It is the simplest and almost unique solution that satisfies 
\(M_X = M_{string} \approx
0.6 \times 10^{18} \) GeV and \( M_{INT}\approx 5\times 10^{11}\) GeV in 
superstring derived SUSY 
SO(10) GUTs.
(ii) The proton is stable enough by the automatic "doublet-triplet splitting" 
closely connected with 
the {\it D}-parity violation. 
(iii) The minimization of SUSY one-loop effective potential in a toy model 
suggests that the SO(10) 
gauge theory tends to break dynamically down to the SU(4) \(\times\) 
SU(2)\(_ L \times\) SU(2)\(_R\) model.

\vglue.5in
\newpage

Supersymmetric (SUSY) SO(10) grand unified theory (GUT) is one of candidates
of true GUT because:

(1) The group SO(10) contains the standard model (SM) 
gauge symmetry SU(3)\(_C \times\) SU(2)\(_L \times\) U(1)\(_Y \). 

(2) All members of each generation of light quarks and leptons belong to 
single irreducible representation {\boldmath $16$} of SO(10) together with a 
right-handed neutrino and
the anomaly cancellation is automatic.

(3) It naturally provides us with an intermediate mass scale, \( M_{INT}\), 
which might develop
a new physics. The case of \( M_{INT}\approx 10^{11-12}\) GeV is of special 
interest in
connection with an invisible axion [1], massive neutrinos through the see-saw 
mechanism [2],
an inflaton for generating the cosmological baryon asymmetry [3] and so on. 

(4) It is a subgroup of E\(_8\) and so the SO(10) GUT could be derived from 
the
compactification of superstring theory (SST) such as E\(_8\times\)  E'\(_8\)
heterotic string theory. 

(5) The grand unification scale \(M_X\) can be raised from O(\(10^{16}\)
GeV) of minimal SUSY standard model (MSSM) to the string scale \(M_{string}
\sim 10^{18}\) GeV.
  
Suppose that the SUSY SO(10) GUT has been somehow derived from an underlying 
SST, e.g., the  E\(_8\times\)  E'\(_8\) heterotic string, by a compactification 
\(\grave a \)
{\it la\/} Witten [4] on some manifold {\it K\/}, and that the situation 
corresponds to level 1
Ka\(\hat{\mbox{c}}\)-Moody algebra representations. Then, the contents of the 
chiral scalar superfields at the GUT 
scale \( M_X\) are given by [4,5]

\begin{equation}
n_g  \mbox{\boldmath $16$} + \delta (\mbox{\boldmath $16$}+ \overline{\mbox
{\boldmath $16$}})
+ \varepsilon \mbox{\boldmath $10$}  + \eta \mbox{\boldmath $1$},
\vspace*{.2in}
\end{equation} 
where \(n_g \) (= 3 ) is the number of generations of light quarks and leptons 
and
\(\delta , \varepsilon\) and \(\eta \) denote the number of Higgs superfields 
of representations \( \mbox{\boldmath $16$} + \overline{\mbox{\boldmath $16$}},
\mbox{\boldmath $10$}\)  and  \(\mbox{\boldmath $1$} \), respectively. The 
values of \(n_g \),
\(\delta , \varepsilon\) and \(\eta \) depend, in principle, on the topology 
of {\it K\/}.

In passing, it should be remarked that the gauge group SO(10) cannot be 
obtained by the simple {\it 
standard embedding} of SU(3) holonomy of {\it K} into E\(_8\) but by an 
embedding of gauge connection of 
holomorphic SU(4) vector bundle over {\it K} [4]. Consequently, the number 
of generations of quarks and 
leptons is not given by one half of Euler characteristics of {\it K} but by 
more general Atiyah-Singer 
index theorem [4,6] which could develop a new potential of explaining \(n_g
\) = 3.

There are several possible paths from the SUSY SO(10) GUT to MSSM. 
For simplicity, we will confine here ourselves to the cases of single 
intermediate 
scale, namely to the breaking pattern:

\begin{equation}
\mbox{SO(10)}\stackrel{M_X}{\longrightarrow}G_{INT}\stackrel{M_{INT}}
{\longrightarrow}\mbox{SU(3)}_C 
\times\mbox{SU(2)}_L \times \mbox{U(1)}_Y,
\vspace*{.2in}
\end{equation}
and investigate the following cases of intermediate symmetries: 
\vspace*{.2in}

\[\quad G_{INT}=\left\{
	  \begin{array}{l}
      \mbox{(a)  SU(4)} \times \mbox{ SU(2)}_L \times \mbox{SU(2)}_R 
\times D, \\
      \mbox{(b)  SU(4)} \times \mbox{ SU(2)}_L \times \mbox{SU(2)}_R, \\
      \mbox{(c)  SU(4)} \times \mbox{ SU(2)}_L \times \mbox{U(1)}_R, \\
      \mbox{(d)  SU(3)}_C \times \mbox{SU(2)}_L \times \mbox{SU(2)}_R 
\times\mbox{U(1)}_{B-L}
 \times D,\\
      \mbox{(e)  SU(3)}_C \times \mbox{SU(2)}_L \times \mbox{SU(2)}_R 
\times\mbox{U(1)}_{B-L},
   \end{array}\right. \] 
where {\it D} in (a) and (d) means that the spectrum of Higgs sector is 
symmetric under
the exchange of \(L\leftrightarrow R\), i.e., {\it D}-parity [7] is
conserved, while in (b) and (e) the {\it D}-parity is violated. We discard 
the case \(G_{INT}=\)SU(5) or
SU(5)\(\times\)U(1) because it requires rather big \(M_{INT}\) which does not 
deserve the name of "{\it
intermediate}".

The purpose of this paper is to show that the path (b) with the absence of the 
component
\((\mbox{\boldmath $4$},\mbox{\boldmath $2$},\mbox{\boldmath $1$})+(\overline
{\mbox{\boldmath $4$}},
\mbox{\boldmath $2$},\mbox{\boldmath $1$})\) in the Higgs multiplet \(\mbox
{\boldmath $16$}
+ \overline{\mbox{\boldmath $16$}}\) has exclusively remarkable virtues.
We call the model with this breaking path a {\it D}-parity violated SUSY 
Pati-Salam model [8]. 
In the following, we present characteristic features of the model in three 
different aspects:
(I) the renormalization group equation (RGE) analysis, (II) the 
"doublet-triplet splitting" by 
Witten mechanism and (III) the minimization of SUSY one-loop effective 
potential in a toy model.

\vspace*{.2in}

\noindent (I) {\it RGE analysis}

In this section, we analyze the evolution of gauge couplings for (2). In all 
the cases (a)\(\sim\)(e), 
we assume that the colored components in the Higgs multiplet {\boldmath $10$} 
are at least as heavy as 
\(M_X\) and decoupled from the theory below the unification scale, so that 
the "doublet-triplet splitting"
has been realized to avoid the danger of fast proton decay. It should be 
noticed here that the colored
Higgses belong to the different representation of \(G_{INT}\) from the 
electroweak doublets. Therefore, 
the "doublet-triplet splitting" could be more easily implemented for 
(a)\(\sim\)(e) than for SU(5). This is
another reason why we do not take SU(5) or SU(5) \(\times\) U(1) as 
candidates of \(G_{INT}\).

We find [9-11] that in the {\it D}-parity violated SUSY Pati-Salam model, 
a simple choice of the Higgs sector, \(2\{(\mbox{\boldmath $4$},\mbox
{\boldmath $1$},\mbox{\boldmath
$2$})+(\overline{\mbox{\boldmath $4$}},\mbox{\boldmath $1$}, \mbox
{\boldmath $2$})\}+
(\mbox{\boldmath $1$},\mbox{\boldmath $2$},\mbox {\boldmath $2$})+\mbox{some}
~
(\mbox{\boldmath $1$},\mbox{\boldmath $1$},\mbox {\boldmath $1$})\)'s which 
corresponds
to \(\delta = 2 \) and \(\varepsilon = 1\) in (1), can attain \(M_X = M_
{string} \approx
0.6 \times 10^{18} \) GeV and, at the same time, \( M_{INT}\approx 5\times 
10^{11}\) GeV [9-11].
The relation \(M_X = M_{string}\) indicates that the intermediate gauge 
symmetry \(G_{INT}\) might be 
directly realized from the string at \(M_{string}\) with the gauge couplings 
unified so that the "GUT" 
group SO(10) need not in fact be embodied explicitly. For such cases we will 
use the word "SO(10) GUT" in the sense 
that \(G_{INT}\subset\) SO(10). 
 
This case of \(G_{INT}\)=SU(4) \(\times\) SU(2)\(_ L \times\) SU(2)\(_R\) with 
the Higgs sector 
\(2\{(\mbox{\boldmath $4$},\mbox{\boldmath $1$},\mbox{\boldmath$2$})+(
\overline{\mbox{\boldmath $4$}},
\mbox{\boldmath $1$}, \mbox{\boldmath $2$})\}+(\mbox{\boldmath $1$},\mbox
{\boldmath $2$},\mbox 
{\boldmath $2$})+\mbox{some}~
(\mbox{\boldmath $1$},\mbox{\boldmath $1$},\mbox {\boldmath $1$})\)'s is 
actually the only possibility that both of the constraints (i) \(M_X = 
M_{string}\) and (ii) \(M_X/M_{INT}\approx 10^6 \) are fulfilled in the SUSY 
SO(10) GUT with the 
single intermediate scale. Indeed, according to the RGE analysis of ref.[12], 
if we demand
the constraints (i) and (ii), the relative changes in the beta functions of 
the MSSM
above \(M_{INT}\) due to the additional Higgs supermultiplets must satisfy 
the condition

\begin{equation}
\frac{2}{5}r\equiv\triangle b_2 - \triangle b_1 = 2, ~~ q \equiv\triangle 
b_3 - \triangle b_2 =1,
\vspace*{.2in}
\end{equation}
where \(b_i = -2\pi \partial \alpha \sp{-1}\sb{i}/\partial\mbox{ln}\mu.\) The 
above case of
{\it D}-parity violated SUSY Pati-Salam model just satisfies (3). In the other 
cases, we have

\begin{eqnarray}
&(a)&~ q+r=18,\\
&(c)&~ r<0,\\
&(d)&~ q+r=0,~9,\\
&(e)&~ q+r=9,~3 \quad \mbox{or} \quad q+r<0,
\end{eqnarray}  
even with the help of exotics. Obviously, (4)\(\sim\)(7) cannot be compatible 
with
the condition (3).

It is noteworthy that, although the cases (a) and (b) are both the SUSY 
Pati-Salam
model, only the case (b) can satisfy (3), namely the {\it D}-parity violation 
is indispensable for
the SUSY Pati-Salam model to realize \(M_X = M_{string}\) and \( M_{INT}
\approx 5\times 10^{11}\) GeV.
In this context, it might not be accidental that, in a simple SST-derived SUSY 
Pati-Salam model [13,10]
constructed by the fermionic formulation [14], the Higgs sector contains two
copies of ({\boldmath $4$},{\boldmath $1$},{\boldmath $2$})+(\(\overline{
\mbox{\boldmath $4$}}\),
{\boldmath $1$},{\boldmath $2$}), but no ({\boldmath $4$},{\boldmath $2$},
{\boldmath $1$})
+(\(\overline{\mbox{\boldmath $4$}}\),{\boldmath $2$},{\boldmath $1$}), so 
that the {\it D}-parity
is violated. In order to make the model more realistic, we must find a 
dynamical mechanism in which
the gauge symmetry breaking \(\mbox{SU(4)} \times \mbox{ SU(2)}_L \times 
\mbox{SU(2)}_R\rightarrow
\mbox{SU(3)}_C\times\mbox{SU(2)}_L \times \mbox{U(1)}_Y\) takes place actually 
at \( M_{INT}\approx
5\times 10^{11}\) GeV. An example of such a mechanism is found in ref.[11].

\vspace*{.2in}

\noindent (II) {\it "Doublet-triplet splitting"}

Hereafter we assume that the compact manifold {\it K} is multiply-connected. 
Then, all fields obey 
nontrivial boundary conditions and the gauge fields can develop vacuum 
expectation values (VEVs) on
{\it K}. The gauge symmetry breaking is dynamically caused by Hosotani 
mechanism [15,16] through the 
modified Wilson loops {\it W}'s[16]. 
Let {\it G} be \(\pi_1(K)\)
and \(\hat G \)  the image of a homomorphim of {\it G} into the gauge group 
SO(10). Then, \(\hat G \)
will be the symmetry of {\it W} and it is only \((G \oplus \hat G \))-invariant 
modes that remain 
massless after the breaking (Witten mechanism [17]).

In the case of the path (b), \(G_{INT}=\) SU(4) \(\times\) SU(2)\(_ L \times\) 
SU(2)\(_R\) with the 
{\it D}-parity violated, the "doublet-triplet splitting" (in this case, the 
decoupling of the component
({\boldmath $6$},{\boldmath $1$},{\boldmath $1$}) in the Higgs multiplet 
{\boldmath $10$}) 
by Witten mechanism is closely connected with the {\it D}-parity violation [9]. 
Namely, it is automatic 
in the sense that any two of the following three statements lead to the 
remaining one (for the proof, 
see ref.[9]):
 
\noindent\quad (A) The SO(10) breaks down to SU(4) \(\times\) SU(2)\(_ L 
\times\) SU(2)\(_R\),
i.e., the component ({\boldmath $6$},{\boldmath $2$},{\boldmath $2$}) in the 
gauge multiplet
{\boldmath $45$} becomes superheavy while the components
({\boldmath $15$},{\boldmath $1$}, {\boldmath $1$}), ({\boldmath $1$},
{\boldmath $3$},
{\boldmath $1$}) and ({\boldmath $1$},{\boldmath $1$}, {\boldmath $3$}) remain 
massless.
 
\noindent\quad(B) The {\it D}-parity is violated, i.e., the component
({\boldmath $4$},{\boldmath $2$},{\boldmath $1$})+(\(\overline{\mbox{\boldmath 
$4$}}\),{\boldmath $2$},
{\boldmath $1$}) in the Higgs multiplet {\boldmath $16$} + \(\overline{\mbox
{\boldmath $16$}}\) becomes 
superheavy while the component ({\boldmath $4$},{\boldmath $1$},{\boldmath $2$}
)+(\(\overline{\mbox
{\boldmath $4$}}\),{\boldmath $1$},{\boldmath $2$}) remains massless, or {\it 
vice versa}.

\noindent\quad(C) The "doublet-triplet splitting" is realized, i.e., the 
component
({\boldmath $6$},{\boldmath $1$},{\boldmath $1$}) in the Higgs multiplet 
{\boldmath $10$} becomes 
superheavy while the component ({\boldmath $1$},{\boldmath $2$}, {\boldmath 
$2$}) remains massless. 
\vspace*{.2in}

From (I) and (II), we understand that, in the {\it D}-parity violated SUSY 
Pati-Salam model,
the achievement of \(M_X = M_{string} \approx 0.5 \times 10^{18} \) GeV and 
the appearance of
\( M_{INT}\approx 5\times 10^{11}\) GeV are closely connected with the 
stability of proton
in terms of the {\it D}-parity violation. As to the potentially dangerous 
Higgs
multiplets ({\boldmath $4$},{\boldmath $1$},{\boldmath $2$})+(\(\overline
{\mbox{\boldmath $4$}}\),
{\boldmath $1$},{\boldmath $2$})'s, which contain color triplet component 
\((\overline{\mbox{\boldmath
$3$}},\mbox{\boldmath $1$})_{2/3}+(\mbox{\boldmath $3$},\mbox{\boldmath $1$}
)_{-2/3}\) of SU(3)\(_C
\times \) U(1)\(_Y\),
their direct couplings with the quarks and leptons are forbidden
by the symmetry of the model. This is due to the fact that {\boldmath $16$}
\(\times\){\boldmath
$16$}\(\times\){\boldmath $16$} or {\boldmath $16$}\(\times\)
{\boldmath $16$}\(\times \overline{\mbox{\boldmath $16$}}\) do not contain 
{\boldmath
$1$} of SO(10) and this feature is inherited by SU(4) \(\times\) SU(2)\(_ L 
\times\) SU(2)\(_R\).
After the breaking of SU(4) to SU(3) at \( M_{INT}\), they effectively couple 
with the quarks
and leptons through the process such as

\begin{eqnarray}
\mbox{\boldmath $16$}+\mbox{\boldmath $16$}\longrightarrow\mbox{\boldmath $10$}
\longrightarrow\mbox
{\boldmath $16$}+<\mbox{\boldmath $16$}>,\nonumber
\vspace*{.2in}
\end{eqnarray}
where \(<~>\) indicates a VEV. However, the strength of induced effective 
coupling {\boldmath $16$}
\(\cdot\){\boldmath$16$}\(\cdot\){\boldmath $16$} is
greatly suppressed by a factor \( M_{INT}/M_X\) and negligible.
\vspace*{.2in}

\noindent (III) {\it SUSY effective potential}

There is a hint that the path of SUSY Pati-Salam model is dynamically 
favorable. It is the embedding 
scheme of \(\hat G \) into SO(10) that characterizes the breaking direction 
[17,18]. However, we cannot 
arbitrarily choose the Wilson loop but  should determine \(\hat G \) through 
the minimization of the
effective potential for the SUSY SO(10) gauge theory. The minimization of SUSY 
one-loop effective 
potential in a toy model suggests that the breaking direction 
SO(10) \(\rightarrow \) SU(4)\(\times\) SU(2)\(_ L \times\) SU(2)\(_R\) or 
SU(4)\(\times\) SU(2)\(_ L 
\times\) SU(2)\(_R\times D(\times D\)
is henceforth abbreviated) is more probable than other directions. Indeed, in 
an SO(10) 
gauge theory on \(M_3 \times S^1\), where \(M_3\) is the three-dimensional 
Minkowski spacetime, the SUSY one-loop effective potential \(V_1\) can be 
estimated by the auxiliary
field tadpole method [19]. The following derivation of \(V_1\) ((16) below) is 
a revised version of 
ref.[20].

We assume that (i) the gauge fields have non-vanishing VEV \(<A_y>\neq 0\)
on \(S^1\), {\it y} indicating the coordinate of \(S^1\), (ii) the VEVs of the 
scalar components of
matter superfields are zero, (iii) the effective potential satisfies the SUSY 
boundary condition,
\(V_1(f=f^+=d=0)=0\) where \(f\equiv<F>\), \(f^+\equiv<F^+>\) and \(d\equiv<D>
\)
denote the VEVs of auxiliary fields and (iv) the gauge symmetry does not break 
spontaneously, which
means that {\it f} is gauge singlet and \(d=0\) so that we need not consider 
the {\it D} tadpoles.

The action is given by [21]

\begin{equation}
S=\int dz[\sum_i\phi_i^+ e^{gV}\phi_i]+[\int d\sigma(\frac14tr(W^\alpha 
W_\alpha)+P(\phi_i))
+\mbox{h.c.}],
\end{equation}
where \(\phi_i\) stands for chiral matter superfields in the irreducible 
representaion
of SO(10) displayed in (1), \(V\equiv V^a T^a\) defines the vector superfield 
with \(T^a\) being 
the generators of SO(10) in the adjoint representaion [22], \(W_\alpha
\equiv -\frac14 \overline D^2 \mbox{exp}(-gV)D_\alpha\mbox{exp}(gV)\) denotes 
the SUSY field strength, 
\(dz=d^3dyd^2\theta d^2\overline \theta\), \(d\sigma = d^3dyd\theta d\overline 
\theta\) and \(P(\phi_i)
=\frac12m_{ij}\phi_i\phi_j+\frac1{3!}\lambda_{ijk}\phi_i\phi_j\phi_k+\cdots\) 
is the superpotential.

By making a component field expansion for (8) in the Wess-Zumino gauge and by 
translating the fields
as \(F_i\rightarrow F_i+f_i\) and \(A_\mu
\rightarrow A_\mu +a\delta_{\mu y}\), where \(A_\mu\) denotes the spin-1 
component of
{\it V} and \(a\equiv<A_y>\), we obtain propagators for spin-0 component 
\(\Phi_i\) of
\(\phi_i\) which are relevant to the {\it F} tadpole as follows:

\begin{equation}
\Phi_i\Phi_j:-\lambda_{ijk}f_k^+/\triangle,
\end{equation}
where
\begin{equation}
\triangle (a, f) =\mbox{det}\{(p_\mu-ga\delta_{\mu y})^4-\lambda_{ijk}
\lambda_{ijk'}f_k^+f_{k'}\}.
\vspace*{.2in}
\end{equation}
In terms of irreducible representaions of SO(10), \(\Phi_i\Phi_j\) and 
\(f_k^+=f_{k'}\equiv f\) in (9)
and (10) in fact belong respectively to {\boldmath $16$}\(\times \overline
{\mbox{\boldmath $16$}}\)
and {\boldmath $1$} due to the assumption (iv). 

Then, using the SUSY boundary condition \(V_1(f=0)=0\), we obtain

\begin{eqnarray}
V_1&=&\frac12\int\frac{d^4p}{(2\pi)^4}[\mbox{ln}\triangle(a,f)-\mbox{ln}
\triangle(a,0)]\nonumber\\
   &=&\frac12\mbox{tr}\int\frac{d^4p}{(2\pi)^4}[\mbox{ln}\{(p_\mu-ga
\delta_{\mu y})^4-\lambda^2f^2\}
-\mbox{ln}\{(p_\mu-ga\delta_{\mu y})^4]\\
   &=&\frac{\lambda^2f^2}{4\pi^2L}\mbox{tr}\sum_{n=-\infty}^\infty \int_0^
\infty p^2dp~\mbox{ln}[p^2
+(\omega_n-ga)^2]+O(f^3),\nonumber
\vspace*{.2in}
\end{eqnarray}
where \(\omega_n=(2\pi n+\beta)/L\), with \(\beta\) being the phase which can 
enter the boundary
conditions on \(S^1\) for the chiral superfields \(\phi_i\)'s and {\it L} 
being the periodicity of
the coordinate {\it y} of \(S^1\). In arriving at the third line of (11), we 
have assumed
\(\vert f\vert \ll L^{-1}\), that is, the SUSY breaking scale is
much smaller than the compactification scale. Under the constraint that 
\(G_{INT}\) has the same rank 
with SO(10) and that SU(3)\(_C \times\) SU(2)\(_L\) is unbroken, the \(16
\times 16\) matrix \(a\) can be 
parametrized in terms of two real free parameters as [16, 18]

\begin{equation}
a=(gL)^{-1}(\theta H^{\theta}+\psi H^{\psi}),
\end{equation}
where
\begin{eqnarray}
\theta H^{\theta}&=&\mbox{diag}(\theta_1,\cdots,\theta_{16})\nonumber\\
&=&\theta\mbox{diag}(-3,-3,1,1,1,1,1,1,3,3,-1,-1,-1,-1,-1,-1),\\
\psi H^{\psi}&=&\mbox{diag}(\psi_1,\cdots,\psi_{16})\nonumber\\
&=&\psi\mbox{diag}(0,0,0,0,0,0,0,0,1,-1,1,1,1,-1,-1,-1).
\vspace*{.2in}
\end{eqnarray}
\vspace*{.2in}
The parametrization (12) corresponds to the Wilson loop
\begin{equation}
W=\mbox{exp}(\theta H^{\theta}+\psi H^{\psi}).
\vspace*{.2in}
\end{equation}
The \(\theta\) and \(\psi\) dependence of \(V_1\) can easily be estimated by 
the method of ref.[23] 
to be

\begin{equation}
V_1(\theta,\psi)=\frac{\lambda^2f^2}{\pi^2L^4}\sum_{j=1}^{16}\sum_{n=-\infty}^
\infty\frac{\mbox{cos}n(\theta_j+\psi_j
-\beta)}{n^4}+(\theta,~\psi-\mbox{independent}), 
\vspace*{.2in}
\end{equation}
where \(O(f^3)-\)term has been neglected. It is evident that, if one takes 
\(\beta=0\), which
means that {\boldmath $16$}- or \(\overline{\mbox{\boldmath $16$}}
\)-dimensional chiral fields obey
a periodic boundary condition, for simplicity, the potential \(V_1\) attains 
the 
minimum at \(\theta=\pi\) and \(\psi=0~(\mbox{mod} 2\pi)\). Unfortunately, the 
corresponding Wilson
loop {\it W} (15) can neither break SO(10) [18] nor determine \(\hat G\). 
Therefore, the minimization of 
\(V_1\) cannot discriminate \(G_{INT}\)'s.

However, it will be reasonable to assume \(\hat G=Z_m\times Z_n\) and take an 
average \(\bar{V}_1(m,n)\) 
of \(V_1(\theta,\psi)\) such as

\begin{equation}
\bar{V}_1(m,n)=\frac{1}{mn-1}\sum_{k=0}^{m-1}\sum_{l=0}^{n-1}\{V_1(\frac{2k\pi}
{m},\frac{2l\pi}{n})
-V_1(0,0)\},
\vspace*{.2in}
\end{equation}
and minimize it in terms of \(m,n\) except for a trivial case \(m=2,n=1\). The 
result is remarkable. 
The \(\bar{V}_1\) attains the minimum when \(\hat G=Z_4~(m=4,n=1)\), which 
gives SO(10) \(\rightarrow 
G_{INT}=\) SU(4)\(\times\) SU(2)\(_L\times\) SU(2)\(_R\). In fact, we obtain 

\begin{equation}
\bar{V}_1(4,1)<\bar{V}_1(2,2)<\bar{V}_1(5,1)<\bar{V}_1(6,1)<\cdots,
\vspace*{.2in}
\end{equation}
which implies that, in the breaking of SO(10) to \(G_{INT}\) via Wilson loops, 
\(G_{INT}=\) SU(4)
\(\times\) SU(2)\(_L\times\) SU(2)\(_R\) by \(\hat G=Z_4\) or \(\hat G=Z_2
\times Z_2\) [18] is more 
probable than \(G_{INT}=\) SU(3)\(_C\times\)SU(2)\(_L\times\)SU(2)\(_R \times\) 
U(1)\(_{B-L}\) as well  
as \(G_{INT}=\) SU(4)\(\times\) SU(2)\(_L\times\) U(1)\(_R\) by \(\hat G=Z_3\), 
\(\hat G=Z_m,~m\geq 5\)
or other \(\hat G\)'s.

It is obvious that the above inference is valid also for \(M_4\times T^6\), 
where \(M_4\) and \(T^6\)
are four-dimensional Minkowski spacetime and six-torus, respectively, which is 
derived from 
E\(_8\times\)  E'\(_8\) heterotic string by torus compactification, provided 
that the VEVs of
gauge fields are isotropic on \(T^6\) in the sense that their 6 components are 
all identical 
independent of the way of their corresponding to \(S^1\)'s in \(T^6\). It will 
not be so trivial
to generalize the above toy model to more realistic cases. However, the 
characteristic feature of 
\(V_1\) (16)
is expected to hold as far as the manifold {\it K} is multiply-connected and 
the VEV of the gauge 
field on {\it K} can be parametrized as (12). It is one of next tasks to 
investigate if such argument
is also applicable to orbifolds or more general multiply-connected compact 
manifolds of dimension 
six than \(T^6\).

In summary, we have seen that the {\it D}-parity violated SUSY Pati-Salam model 
with the Higgs sector 
consisting of \(2\{(\mbox{\boldmath $4$},\mbox{\boldmath $1$},\mbox{\boldmath
$2$})+(\overline{\mbox{\boldmath $4$}},\mbox{\boldmath $1$}, \mbox{\boldmath 
$2$})\}+
(\mbox{\boldmath $1$},\mbox{\boldmath $2$},\mbox {\boldmath $2$})+\mbox{some}~
(\mbox{\boldmath $1$},\mbox{\boldmath $1$},\mbox {\boldmath $1$})\)'s has the 
following advantages: (i)
It is the simplest and almost unique solution that satisfies \(M_X = M_{string}
 \approx
0.6 \times 10^{18} \) GeV and \( M_{INT}\approx 5\times 10^{11}\) GeV in 
SST-derived SUSY SO(10) GUTs.
(ii) The stability of proton is guaranteed by the automatic "doublet-triplet 
splitting" and closely 
connected with the {\it D}-parity violation which is indispensable for the 
advantage (i) to be realized. 
(iii) The minimization of SUSY one-loop effective potential in a toy model 
suggests that the SUSY SO(10) 
gauge theory is expected to break dynamically down to the SUSY Pati-Salam 
model.
\vspace*{.2in}
\newpage
\newcounter{0000}
\begin{list}
{[~\arabic{0000}~]}{\usecounter{0000}
\labelwidth=0.8cm\labelsep=.05cm\setlength{\leftmargin=0.7cm}
{\rightmargin=.2cm}}
\item J.E.~Kim, Phys.~Rep.~{\bf 149}~(1987)~1.
\item T.~Yanagida, in Proc. Workshop on Unified Theory and the Baryon Number 
in the Universe,
eds. O.~Sawada and A.~Sugamoto, KEK Report No.79-18(1979), p.95; M.~Gell-Mann, 
P.~Ramond and R.~Slansky, 
in Supergravity, eds. P.van Niewwenhuizen and D.~Freedman (North-Holland, 
Amsterdam, 1979), p.315. 
\item M.~Fukugita and T.~Yanagida, Phys.~Lett.~{\bf 174}~(1986)~45;

B.A.~Campbell, S.~Davidson and K.A.~Olive, Nucl.~Phys.~{\bf B399}~(1993)~111.
\item E.~Witten, Nucl.~Phys.~{\bf B268}~(1986)~79.
\item R.~Holman and D.B.~Reiss, Phys.~Lett.~{\bf 176}~(1986)~74; 

D.~Bailin, A.~Love and S.~Thomas, Phys.~Lett.~{\bf 176}~(1986)~81.
\item I.~Bars and M.~Visser, Phys.~Lett.~{\bf B163}~(1985)~118;

M.J.~Hayashi, A.~Murayama and S.~Takeshita, Phys.~Lett.~{\bf B228}~(1989)~42;
~ibid.~{\bf B247} ~(1990)~525;

A.~Murayama, Phys.~Lett.~{\bf B248}~(1990)~277.
\item D.~Chang, R.N.~Mohapatra and M.K.~Parida, Phys.~Rev.~Lett.~{\bf 52}~
(1984)~1072.
\item J.C.~Pati and A.~Salam, Phys.~Rev.~Lett.~{\bf 31}~(1973)~661;~Phys.~Rev.
~{\bf D10}~(1974)~275.
\item H.M.~Asatryan and A.~Murayama, Int.~J.~Mod.~Phys.~{\bf A7}~(1992)~5005.
\item A.~Murayama and A.~Toon, Phys.~Lett.~{\bf B318}~(1993)~298.
\item A.~Murayama, Int.~J.~Mod.~Phys.~{\bf A12}~(1997)~903;~Phys.~Lett.~{\bf 
B324}~(1994)~366; 
{\it Erratum}~-ibid.~{\bf B329}~(1994)~526.
\item S.P.~Martin and P.~Ramond, Phys.~Rev.~{\bf D51}~(1995)~6515.
\item I.~Antoniadis and G.K.~Leontaris, Phys.~Lett.~{\bf B216}~(1989)~333;

I.~Antoniadis, G.K.~Leontaris and J.~Rizos, Phys.~Lett.~{\bf B245}~(1990)~161.
\item H.~Kawai, D.C.~Lewellen and S.-H.H.~Tye, Phys.~Rev.~Lett.~{\bf 57}~
(1986)~1832;~Phys.~Rev. ~
{\bf D34}~ (1986)~3794;~Nucl.~Phys.~{\bf B288}~(1987)~1;

I.~Antoniadis, C.P.~Bachas and C.~Kounnas, Nucl.~Phys.~{\bf B289}~(1987)~87;

I.~Antoniadis and C.~Bachas,  Nucl.~Phys.~{\bf B298}~(1988)~586.
\item Y.~Hosotani, Phys.~Lett.~{\bf B126}~(1983)~309.
\item Y.~Hosotani, Ann.~Phys.~(NY)~{\bf 190}~(1989)~233.
\item E.~Witten, Nucl.~Phys.~{\bf B258}~(1985)~75.
\item B.R.~Greene, K.H.~Kirklin and P.J.~Miron, Nucl.~Phys.~{\bf B274}~(1986)
~574.
\item R.D.C.~Miller, Phys.~Lett.~{\bf B124}~(1983)~59;~Nucl.~Phys.~{\bf B229}
~(1983)~189. 
\item A.~Murayama, Phys.~Lett.~{\bf B267}~(1991)~88.
\item J.~Wess and J.~Bagger, Supersymmetry and supergravity (Princeton U.P., 
Princeton, NJ, 1983).
\item S.~Rajpoot, Phys.~Rev.~{\bf D22}~(1980)~2244.
\item D.J.~Gross, R.D.~Pisarski and L.G.~Yaffe, Rev.~Mod.~Phys.~{\bf 53}~
(1981)~43.
\end{list}

\end{document}